\documentclass[10pt,conference]{IEEEtran}
\IEEEoverridecommandlockouts

\usepackage{hyperref}
\usepackage{cite}
\usepackage{url}
\usepackage{multirow}
\usepackage{amsmath,amssymb,amsfonts}
\usepackage{algorithmic}
\usepackage{graphicx}
\usepackage{textcomp}
\usepackage{xcolor}
\usepackage{listings}
\usepackage{enumitem}
\usepackage{setspace}
\usepackage{amsmath}
\usepackage{amssymb}
\usepackage{framed}
\usepackage{color}
\usepackage{marvosym}
\usepackage{makecell}
\definecolor{light-gray}{gray}{0.85}
 \newcommand{\tabincell}[2]{\begin{tabular}{@{}#1@{}}#2\end{tabular}}
\def\BibTeX{{\rm B\kern-.05em{\sc i\kern-.025em b}\kern-.08em
    T\kern-.1667em\lower.7ex\hbox{E}\kern-.125emX}}

\begin{document}

\title{\textcolor{black}{Extracting Concise Bug-Fixing Patches from Human-Written Patches in Version Control Systems}}

\author{
	\IEEEauthorblockN{Yanjie Jiang$^1$, Hui Liu$^1$$^*$\thanks{$*$ Corresponding author: Hui Liu}, Nan Niu$^2$, Lu Zhang$^3$, Yamin Hu$^1$}
	$^1$School of Computer Science and Technology, Beijing Institute of Technology, China,\\
	$^2$Department of Electrical Engineering and Computer Science, University of Cincinnati, USA,\\
	$^3$Key Laboratory of High Confidence Software Technologies, Peking University, China\\
	Email: \{yanjiejiang,liuhui08,ymhu\}@bit.edu.cn, nan.niu@uc.edu, zhanglu@sei.pku.edu.cn}

\maketitle
\vspace{-0.4cm}
\begin{abstract}
High-quality and large-scale repositories of real bugs and their concise patches collected from real-world applications are critical for research in software engineering community. \textcolor{black}{In such a repository, each real bug is explicitly associated with its fix.} Therefore, on one side, \textcolor{black}{the real bugs and their fixes} may inspire novel approaches for finding, locating, and repairing software bugs; on the other side, \textcolor{black}{the real bugs and their fixes} are indispensable for rigorous and meaningful evaluation of approaches for software testing, fault localization, and program repair. To this end, a number of such repositories, e.g., Defects4J, have been proposed. However, such repositories are rather small because their construction involves expensive human intervention. Although bug-fixing code commits as well as associated test cases could be retrieved from version control systems automatically, existing approaches could not yet automatically extract concise bug-fixing patches from  bug-fixing commits because such commits often involve bug-irrelevant changes. In this paper, we propose an automatic approach, called \emph{BugBuilder}, \textcolor{black}{to extracting complete and concise bug-fixing patches from human-written patches in version control systems.
It excludes refactorings by detecting refactorings involved in bug-fixing commits, and reapplying detected refactorings on the faulty version. It enumerates all subsets of the remaining part and validates them on test cases.  If none of the subsets has the potential to be a complete bug-fixing patch, the remaining part as a whole is taken as a complete and concise bug-fixing patch.}
Evaluation results on 809 real bug-fixing commits in Defects4J suggest that BugBuilder successfully generated complete and concise \textcolor{black}{bug-fixing} patches for forty percent of the bug-fixing commits, and its precision (99\%) was even higher than human experts.
\end{abstract}


\section{Introduction}
\label{section:Introduction}
High-quality and large-scale repositories of real bugs and their concise patches collected from real-world applications are critical for research in  software engineering community.
On one side, such real bugs/patches are indispensable for rigorous evaluation of numerous
automatic or semi-automatic approaches to identifying faulty software applications~\cite{xia2016hydra,majd2020sldeep,he2013learning,nam2019bug}, to locating faulty statements~\cite{wong2016survey,sohn2017fluccs,li2019deepfl,pearson2017evaluating,lee2018bench4bl}, and to repairing faulty applications~\cite{jeffrey2009bugfix,martinez2017automatic,tan2015relifix,daniel2009reassert}. Such approaches are expected to work on real-world applications. Consequently,
it is critical to evaluate such approaches with a large number of real bugs/patches from real-world applications before they could be widely applied in the wild~\cite{just2014defects4j}. On the other side, real bugs and patches may also inspire novel ideas in finding, locating, and repairing software bugs. For example, by analyzing real bugs, researchers could identify what kind of statements are more error-prone and thus try to repair such statements first during automatic program repair~\cite{NOPOL}. Another typical example is the common fix patterns learned from human-written patches~\cite{HumanLearningICSE13}. Leveraging such patterns significantly increased the performance of automatic program repair~\cite{HumanLearningICSE13}. Finally, data-driven and learning-based approaches in automatic program repair~\cite{Learning2RepairTSE} and bug detection~\cite{Leaning2DetectTSE} usually depend on a large number of diverse real bugs/patches.

Bug repositories have been proposed~\cite{just2014defects4j} to facilitate bug-related research. The first category of bug repositories is constructed manually. Typical examples of this category include SIR~\cite{do2005supporting}, BugBench~\cite{lu2005bugbench}, IntroClass~\cite{le2015manybugs}, Codeflaws~\cite{tan2017codeflaws}, QuixBugs~\cite{lin2017quixbugs}, DroixBench~\cite{tan2018repairing}, and DBGBench~\cite{bohme2017bug}. These repositories are constructed by hand, and thus are quite limited in scale and diversity. The second category of bug repositories is constructed automatically or semi-automatically.  iBUGS\cite{dallmeier2007extraction} and ManyBugs~\cite{le2015manybugs} are typical bug repositories constructed by automatically extracting bug-fixing commits as patches. However, existing studies~\cite{just2014defects4j} suggest that bug-fixing commits often contain both bug-fixing changes and bug irrelevant changes, e.g., refactorings. As a result, it is risky to take all of the changes in a bug-fixing code commit as the patch of the associated bug report: The resulting patch may contain code changes irrelevant to the bug.
\textcolor{black}{Code changes are deemed as bug-irrelevant if they are changing/adding/removing functionalities that are not associated with the bug report or they are function-irrelevant common refactorings that could be conducted independently before bug-fixing changes are made. A bug-fixing patch extracted from a bug-fixing commit is deemed as complete and concise if and only if the patch is composed of all bug-relevant changes (complete) but no bug-irrelevant changes are included (concise).}
To extract complete and concise bug-fixing patches, Defects4J~\cite{just2014defects4j}  takes all changes in a bug-fixing commit as a patch, and then manually excludes bug-irrelevant changes from the generated patch. As a result, the resulting patches in Defects4J are highly accurate, often both complete and concise, but the scale and diversity of the patches remain limited. \textcolor{black}{ Notably, applying the bug-irrelevant changes to the original buggy version (called $V_{n-1}$) results in a new version called $V_{bug}$~\cite{just2014defects4j}, and applying the concise patch in Defects4J to $V_{bug}$ fixes the bug and results in the fixed version (called $V_n$).}

Although bug-fixing commits could be identified automatically by comparing bug IDs in bug tracking systems and code commit messages in version control systems~\cite{saha2018bugs}, it remains challenging to extract automatically bug-fixing changes (i.e., patches) from bug-fixing commits.
\textcolor{black}{To fully automate the construction of bug-patch repositories, in this paper, we propose an automatic approach, called \emph{BugBuilder}, to extracting concise  bug-fixing patches from human-written patches in version control systems. It first leverages refactoring mining and reapplication to remove refactorings from human-written patches. It validates whether the remaining part is a complete and concise bug-fixing patch by enumerating all subsets of the remaining part and validating them on test cases. If none of the subsets can be a complete patch, the remaining part as a whole is deemed as a complete and concise patch. Consequently, if the human-written patch is composed of both refactorings and bug-fixing changes, BugBuilder splits it into two ordered patches: a refactoring patch and a following bug-fixing patch. It is highly similar to Defects4J that splits an existing patch into a bug-irrelevant patch and a following bug-fixing patch. Notably, directly applying the bug-fixing patch to the original buggy version would not work. Consequently, to evaluate automated program repair tools/algorithms (called APR tools for short) with the extracted bug-fixing patches, we should first apply the refactoring patch to the original buggy version (called $V_{n-1}$), take the resulting version (called $V_{n-1}'$) as the buggy program to be fixed, generate patches for  $V_{n-1}'$ with the APR tools, and compare the generated patches against the  bug-fixing patch extracted by BugBuilder. Besides APR tools, fault localization tools/algorithms may also leverage the patches extracted by BugBuilder for quantitative evaluation (taking $V_{n-1}'$ as the buggy program to be fixed).} 

BugBuilder has been evaluated on 809 real bug-fixing commits collected by Defects4J. On each of the evaluated commits, we leveraged BugBuilder to extract concise patches automatically. If a patch was successfully generated, we compared it against the manually constructed patch provided by Defects4J. On 809 bug-fixing commits, BugBuilder  automatically generated 324 patches where 308 were identical to manually constructed patches in Defects4J. For the other 16 patches that were different from manually constructed patches, we manually analyzed the associated bug reports as well as the code commits. Evaluation results suggest that out of the 16 pairs of mismatched patches,  12 were caused by incomplete patches in Defects4J whereas the generated patches were complete and concise. Only four out of the 324 generated patches were inaccurate (complete but not concise), and all of them were caused by incomplete detection of refactorings.

The paper makes the following contributions:
\begin{itemize}[listparindent=-0.5cm,leftmargin=0.3cm,topsep=0.06cm]
  \item First, we propose an automatic approach to \textcolor{black}{extracting complete and concise bug-fixing patches from human-written patches in version control systems.} To the best of our knowledge, it is the first fully automatic approach for this purpose. The approach makes it practical to \textcolor{black}{automatically build large-scale high-quality bug-patch repositories, which may significantly facilitate future bug-related research, especially automated program repair and fault localization.}
  \item Second, we evaluate the proposed approach on real bug-fixing code commits. It successfully extracted complete and concise bug-fixing patches for 40\% of the  bug-fixing commits with a high precision of 99\%. The replication package, including the source code of BugBuilder, is publicly available at~\cite{Patches}.
\end{itemize}
\section{Related Work}
\label{section:RelatedWork}
Because of the importance of real bugs/patches, a number of bug repositories have been proposed.
To the best of our knowledge, the software-artifact infrastructure repository (SIR)~\cite{do2005supporting} is the first attempt to provide a database of bugs. It consists of 17 C programs and 7 Java programs. Each of the programs has several different versions together with a set of known bugs and  test suites. However, most of the programs  are very small, and the bugs mostly are hand-seeded or obtained from mutation.
Spacco et al.~\cite{spacco2005software} collected real bugs made by students during programming tasks. Their bug repository contains hundreds of faulty projects accompanied with test cases. However, most of the collected student projects are small, and they could be significantly different from real-world software applications in  industry. IntroClass~\cite{le2015manybugs} proposed by Le et al., Codeflaws~\cite{tan2017codeflaws} proposed by Tan et al., and
QuixBugs~\cite{lin2017quixbugs} proposed by Lin et al. contain real bugs from programming competitions/challenges. However, such bugs made in programming assignments or competitions could be significantly different from real bugs in industry. To this end, Lu et al.~\cite{lu2005bugbench}  manually collected 19 real bugs from 17 programs.
Besides that, Tan et al.~\cite{tan2018repairing} manually collected 24 reproducible crashes from 15 open-source Android Apps, and  B{\"o}hme et al.~\cite{bohme2017bug} requested twelve experts to collect 27 real bugs from open-source C projects. \textcolor{black}{BugsJS~\cite{gyimesi2019bugsjs} is composed of 453 real manually validated JavaScript bugs from JavaScript server-side programs.}

Manual collection of real bugs is tedious and time-consuming. Consequently, automatic and semi-automatic approaches have been proposed to collect real bugs. To the best of our knowledge, iBUGS\cite{dallmeier2007extraction} is a 
semi-automatic approach to collecting real bugs. It extracted 369 bugs from version control systems automatically, assuming that all changes in the bug-fixing commits are bug-related. However, the existing study~\cite{just2014defects4j} suggests that the assumption is not always true, and bug-fixing commits often contain bug-irrelevant changes like refactorings and implementation of new features. Consequently, taking all changes in bug-fixing commits may result in \emph{unconcise patches} that contain both bug-fixing changes and bug-irrelevant changes.
Similar to iBUGS, ManyBugs~\cite{le2015manybugs} also takes the whole bug-fixing commit as a patch and does not exclude irrelevant changes within the commit.
To exclude such bug-irrelevant changes, Defects4J~\cite{just2014defects4j} requests manual isolation of bug-fixing changes from other bug-irrelevant changes.  As a result of the manual isolation, bugs and patches in Defects4J are highly accurate. Consequently, Defects4J becomes one of the most frequently used bug repositories in the community. Another significant advantage of Defects4J is that it also provides an extensible framework to enable reproducible studies in software testing research. However, the manual intervention requested by Defects4J prevents it from being fully automatic, and thus the dataset remains limited in scale and diversity. Bugs.jar~\cite{saha2018bugs} is another large bug repository, containing 1,158 real bugs and patches collected from open-source applications (Apache@GitHub). Similar to Defects4J, it locates bug-fixing commits
by comparing the ID of bug-fixing issues in Jira (a popular
issue tracking systems) and commit messages in GitHub (source code version control system). It differs from Defects4J in that it requests experts to manually verify that the involved bug reports are real bugs, i.e., they are not misclassified as bugs in the Jira repository.

%

BEARS proposed by Madeiral et al.~\cite{madeiral2019bears} collects bugs based on Continuous Integration (CI) instead of source code version control systems. The core step of its bug collection is the execution of the test suite of the program on two consecutive versions. If a test failure is found in the faulty version  and no test failure is found in its patched version, BEARS takes the two versions as the faulty and the fixed versions whereas their difference is taken as the associated patch. However, BEARS does not distinguish bug-fixing changes from bug-irrelevant changes within the same source code commit.
Similar to BEARS, BugSwarm~\cite{tomassi2019bugswarm} also collects bugs/patches from CI and fails to exclude bug-irrelevant changes from the collected patches as well.

We conclude based on the preceding analysis that existing bug repositories are \textcolor{black}{limited in} either scale or quality. Manually or semi-automatically constructed repositories are \textcolor{black}{limited in} scale because they request expensive human intervention. In contrast, automatically constructed ones are \textcolor{black}{limited in} quality because the automatically extracted patches often contains bug-irrelevant changes.
To this end, in this paper, we propose a fully automatic approach to extracting high-quality patches that are both complete and concise.

\textcolor{black}{DiffCat proposed by Kawrykow and Robillard~\cite{kawrykow2011non} identifies non-essential changes (especially refactorings) in version histories, highly similar to RefactoringMiner~\cite{RefactoringMinerICSE} that our approach leverages to identify refactorings. However, at best it may serve as only the first step for concise patch generation (see Fig~\ref{fig:Overview} for more details). Simply recommending refactoring-excluded patches (output of DiffCat) would result in numerous unconcise patches when commits contain non-refactoring bug-irrelevant changes. As a result, developers have to manually check/clean all of the recommended patches to guarantee the quality.
Thung et al.~\cite{thung2013automatic} identify root causes of bugs, i.e., \emph{lines of code in the buggy version that is responsible for the bug}. Such root causes are essentially different from concise patches. Consequently, these approaches~\cite{kawrykow2011non,thung2013automatic} do not address the same issue (automatically constructing bug-patch repositories) as we do. Notably, neither of them leverages off-the-shelf refactoring mining tools and neither of them reapplies discovered refactorings as we do.}

\section{Motivating Example}
\label{section:MotivatingExamples}
\begin{figure}
	\center
	\includegraphics[width=90mm]{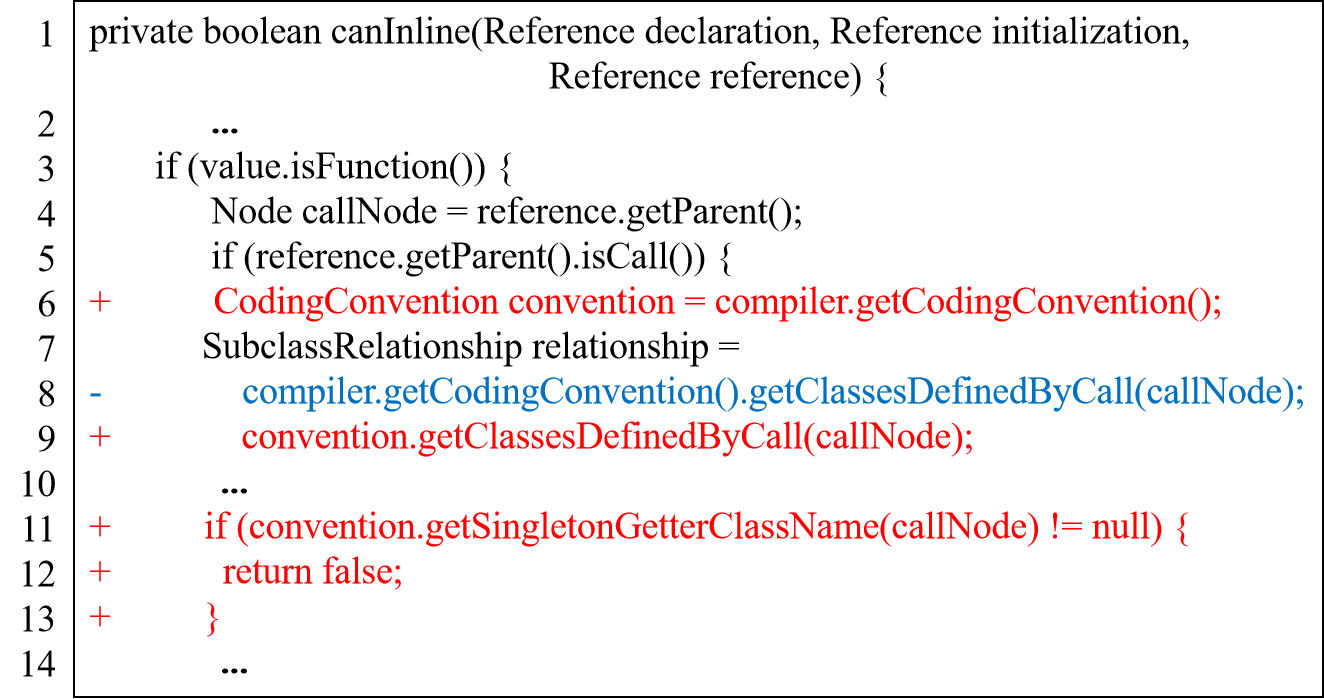}
	\caption{A Bug-Fixing Commit from Google Closure Compiler}
	\label{fig:Sample-refactoring1}
\end{figure}

In this section, we illustrate the challenges in the automatic extraction of concise patches from bug-fixing commits. The example is presented in Fig.~\ref{fig:Sample-refactoring1}, where the changes involved in the commit are highlighted in standard \emph{diff} style. Lines beginning with `-' are removed by the commit whereas lines beginning with `+' are inserted by the commit. Other lines are untouched.
This bug-fixing source code commit comes from Google Closure Compiler~\cite{closure}, and the associated bug report is publicly available online~\cite{closurebug}.


The modified method \emph{canInline} (Line 1) checks whether the provided reference and declaration can be safely inlined.
The bug report~\cite{closurebug}  complains that Singleton getters are inlined although such methods should not be inlined. To fix this bug, developers inserted an  \emph{if} statement (Lines 11-13) that declares Singleton getters could not be inlined (i.e., forcing the method to return \emph{false}). Such changes constitute the bug's complete and concise patch~\cite{just2014defects4j} in Defects4J.

However, we also notice that the bug-fixing commit made more changes than the insertion of an \emph{if} statement (i.e., the complete and concise patch for the bug). Changes on Lines 6, 8, and 9 are bug-irrelevant. Such changes are in fact a typical \emph{extract variable} refactoring. Consequently, extracting  all changes in the bug-fixing commit as the patch would result in unconcise patch that contains bug-irrelevant changes.


BugBuilder successfully generates the complete and concise patch from the bug-fixing commit as follows.
First, it identifies the \emph{extract variable} refactoring by analyzing the changes made in the commit. Second, it reapplies the identified refactoring to the faulty version, which results in a new version $V_{n-1}'$, called \emph{refactoring-included} version. Third it computes all changes (noted as $Chs$)  between  $V_{n-1}'$ and the fixed version ($V_{n}$). Notably, $Chs$ does not include the \emph{extract variable} refactoring because the variable \emph{convention} (Line 6) is defined  in both $V_{n-1}'$ and $V_{n}$. $Chs$ is composed of the changes on Lines 11-13 only. Fourth, BugBuilder enumerates and validates all possible subsets of $Chs$. However, applying any proper subset of $Chs$ to $V_{n-1}'$ results in compiler errors or fails to pass any new test cases in the fixed version. Consequently, such proper subsets could not be taken as candidate patches. Finally, BugBuilder applies all of the changes together (i.e., $Chs$) to $V_{n-1}'$,  resulting in a legal version that passes all test cases associated with the faulty version or the fixed version. Consequently, all of the changes together make up the only candidate patch, and thus  BugBuilder outputs it as the patch for the associated bug.

\section{Approach}
\label{section:Approach}
\subsection{Overview}
\label{sub:Overview}
\begin{figure}
	\center
	\includegraphics[width=85mm]{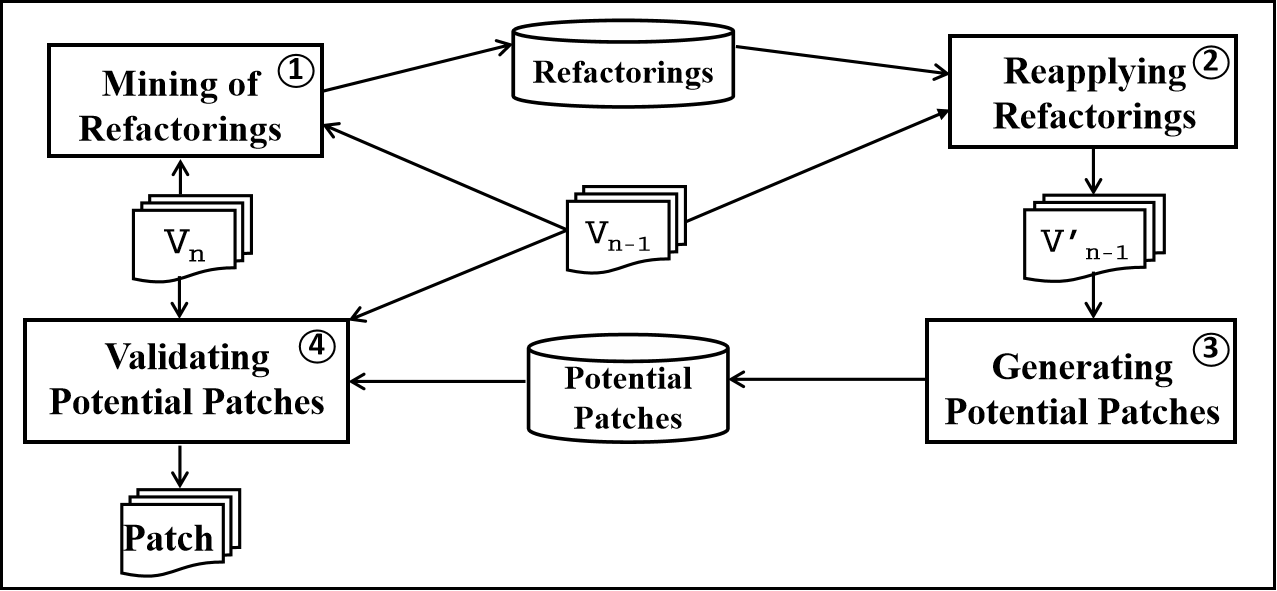}
	\caption{Overview of BugBuilder}
	\label{fig:Overview}
\vspace{-0.4cm}
\end{figure}

An overview of the proposed approach is presented in Fig.~\ref{fig:Overview}. It takes as input two consecutive versions of a software application, i.e., $V_{n-1}$ and $V_n$.  The latter version $V_n$ is called $V_{fix}$ or \emph{fixed version} in Defects4J~\cite{just2014defects4j}, and $V_{n-1}$ is called \emph{faulty version}.  Notably, the two consecutive versions are accompanied by their test cases, noted as $T_{n-1}$ and $T_n$, respectively. $T_{n-1}$ and $T_n$ exclude such test cases that fail on their associated version of the application. The evolution from $V_{n-1}$ to $V_{n}$ is driven by a bug-fixing commit whose commit message contains an ID of a validated bug report.
With such input, the proposed approach BugBuilder works as follows:
\begin{itemize}[listparindent=-0.5cm,leftmargin=0.3cm,topsep=0.06cm]
  \item First, it leverages \emph{RefactoringMiner}~\cite{RefactoringMinerICSE} to discover refactorings that have been applied to $V_{n-1}$. \emph{RefactoringMiner} is a state-of-the-art approach to mining software refactorings by comparing two consecutive versions of the same application. It would result in a list of refactorings, noted as $R$.
  \item Second, if any refactorings have been discovered in the preceding step, i.e., $R$ is not empty, BugBuilder leverages refactoring APIs to apply all of the discovered refactorings to $V_{n-1}$. Applications of such refactorings on $V_{n-1}$ results in a new version $V_{n-1}'$ that is different from both $V_{n-1}$ and $V_n$. For convenience, we  call it \emph{refactoring-included version}.
  \item Third, BugBuilder computes the difference between  the \emph{refactoring-included version} ($V_{n-1}'$) and the fixed version $V_n$. The difference is represented as a sequence of changes (e.g., removing or inserting a token), noted as $Chgs$.
  \item Fourth, BugBuilder enumerates all possible subsequences of $Chgs$, and validates whether the subsequence represents a candidate patch. To validate a subsequence $schg\subseteq Chgs$, BugBuilder applies all changes in $schg$ to $V_{n-1}'$, which results in a new version $V_{n-1}''$. $schg$ represents a candidate patch if and only if   $V_{n-1}''$ passes all test cases in $T_{n-1}$ and passes some test cases in $T_n$ that fail on $V_{n-1}$.
  \item Finally, if only a single candidate patch is generated by BugBuilder,  BugBuilder outputs it as a patch for the associated bug report. Otherwise, no patch would be outputted.
\end{itemize}

\textcolor{black}{Notably, the patches generated by BugBuilder are intended to be applied to $V_{n-1}'$ (refactoring-included version) instead of $V_{n-1}$ (the original buggy version). Consequently, applying such patches to the original buggy version may result in compilation errors and may fail to fix the defects. Although we may revise BugBuilder to generate patches that could be directly applied to the original buggy version, we decide to follow the widely used Defects4J whose patches are also intended to be applied to $V_{n-1}'$ (called $V_{bug}$ in Defects4J~\cite{just2014defects4j})  so that existing tools and algorithms could seamlessly switch from Defects4J to ours.}
\subsection{Detecting and Reapplying Refactorings}
\label{sub:Refactorings}
The key to exclude refactorings from bug-fixing patches is to discover refactorings involved in bug-fixing commits and remove such refactorings before patches are generated.
A few automatic approaches\cite{weissgerber2006identifying,silva2017refdiff,kim2011empirical,palomba2017exploratory,dig2008effective} have been proposed to discover refactorings from version control systems for various reasons, e.g., to facilitate the evaluation of automatic refactoring recommendation algorithms, empirical studies on code evolution, and library API migration. However, to the best of our knowledge, such approaches have not yet been applied to automatic extraction of patches.

A brief introduction to automatic refactoring detection is presented as follows to make the paper self-contained, and more details are referred to related work~\cite{RefactoringMinerICSE,RefactoringMinerTSE}. An automatic refactoring detection algorithm takes as input two consecutive versions (noted as $V_{n-1}$ and $V_n$, respectively) of the same application, and matches elements (e.g., classes, methods, and variables) across versions. Based on the matched elements, the algorithm identifies which elements in the former version (i,e., $V_{n-1}$) have been removed, which elements in the latter version (i.e., $V_n$) have been added, and which elements are kept untouched. The algorithm then defines a list of rules to detect refactorings based on the removed, added, and untouched elements. For example, if a method $m$ in class $C_{1}$ (of version $V_{n-1}$) matches a method $m'$ in class $C_{2}$ (of version $V_n$) and $C_{1}$ does not match $C_{2}$, the algorithm recognizes the changes as a \emph{move method} refactoring that moved method  $m$ from class $C_{1}$  to class $C_{2}$. The performance of the algorithm depends on the accuracy of the employed element matching algorithms as well as the quality of the employed heuristic rules. \emph{RefactoringMiner}~\cite{RefactoringMinerICSE} is more accurate than the alternative algorithms because it leverages an AST-based statement matching algorithm that does not require any user-defined thresholds~\cite{RefactoringMinerTSE}.  To this end, we leverage \emph{RefactoringMiner} to discover refactorings in bug-fixing code commits (excluding those on test cases).

The proposed approach excludes the discovered refactorings by reapplying such refactorings on the faulty version $V_{n-1}$, and employs the resulting version (called $V_{n-1}'$) instead of the original faulty version $V_{n-1}$ to generate patches.  The rationale is that we can divide the revision (bug-fixing commit) into two steps: (\romannumeral1) applying refactorings on $V_{n-1}$, which results in an intermediate version $V_{n-1}'$; and (\romannumeral2) fixing bugs and implementing new features (if there is any) on $V_{n-1}'$. For convenience, we  call the intermediate version $V_{n-1}'$  \emph{refactoring-included version}. Notably, reapplication of the discovered refactorings is accomplished by calling Eclipse refactoring APIs\cite{eclipseRefactoring}. Such APIs are widely used and well-established. For example, if a bug-fixing commit contains a \emph{rename} refactoring that changes the name of method  $m$  from \emph{oldName} to \emph{newName}, we reapply the refactoring as follows:%
\lstset{language=java,
	breaklines,
	xleftmargin=1.3em,xrightmargin=0em,	
	basicstyle=\bfseries\fontsize{6}{6.5}\selectfont\ttfamily,
	escapeinside=``,}
\begin{spacing}{1.4}
\begin{small}
\begin{lstlisting}[numbers=left, frame=single,escapechar=!]
!\underline{rename(m,newName);}!
public void rename (IJavaElement element, String newName){
    ...
  !\colorbox{light-gray}{RenameJavaElementDescriptor descriptor =}!
                   !\colorbox{light-gray}{createRenameDescriptor(element, newName);}!
  !\colorbox{light-gray}{RenameSupport renameSupport=RenameSupport.create(descriptor);}!
   Shell shell=PlatformUI.getWorkbench().getActiveWorkbenchWindow().getShell();
  !\colorbox{light-gray}{renameSupport.perform(shell, PlatformUI.getWorkbench()}!
      !\colorbox{light-gray}{.getActiveWorkbenchWindow());}!
\end{lstlisting}
	\end{small}
\end{spacing} Invocations of Eclipse refactoring APIs are highlighted with gray background in the code snippet.
Notably, we have to customize the code snippet for different categories of refactorings to forward refactoring  information from \emph{RefactoringMiner} to refactoring APIs because the required refactoring information and refactoring APIs vary significantly among different categories of refactorings. Currently, we have customized the code snippet for  eight most common refactorings, including \emph{rename classes}, \emph{rename method}, \emph{rename variables}, \emph{rename fields}, \emph{rename parameters}, \emph{rename packages}, \emph{extract methods}, and \emph{extract variables}. Such refactorings account for 72\%=(284/397) of the refactorings discovered in the bug-fixing commits in Defects4J.

\subsection{Generating Potential Patches}
\label{sub:PatchGeneration}
For a given intermediate version $V_{n-1}'$ and the bug-fixing version $V_n$, the proposed approach generates all possible patches. To this end, it  computes the difference between $V_{n-1}'$ and  $V_n$ (excluding their differences in test cases). The difference is represented as a sequence of token-level changes (e.g., removing or inserting a token), noted as $Chgs=<chg_1, chg_2,\dots , chg_k>$. Each of the token-level changes is composed of three parts: position, token, and edition type. Edition type is either \emph{remove} or \emph{insert}.

The proposed approach enumerates all subsequences of $Chgs$. Each subsequence $schg \subseteq Chgs$ represents a potential patch that makes all of the token-level changes in $schg$ on  $V_{n-1}'$, and ignores other changes in $Chgs$. To reduce the number of potential patches, we also introduce coarse-grained changes: line-level changes. If a whole line of source code has been removed from $V_{n-1}'$, we represent it as a line-level change instead of a sequence of token-level changes. Insertion of a new line of source code is handled in the same way as a line-level change. Consequently, a potential patch is finally represented as a sequence of token-level and/or line-level changes.

\subsection{Validating Potential Patches}
\label{sub:patchValidation}
The validation of a potential patch $pt$ is conducted in two phases. In the first phase, the proposed approach applies this potential patch to the intermediate version $V_{n-1}'$,  resulting in a new version $V_{n-1}''$. If $V_{n-1}''$ could not  compile successfully, the potential patch is discarded as an illegal patch and its validation terminates. In the second phase, $V_{n-1}''$  is further validated with test cases associated with the faulty version (noted as $T_{n-1}$) and those associated with the fixed version (noted as $T_{n}$) as follows:
\begin{itemize}[listparindent=-0.5cm,leftmargin=0.3cm,topsep=0.06cm]
  \item If any test case in $T_{n-1}$  fails on $V_{n-1}''$, $pt$ is not a valid patch and  its validation terminates;
  \item  The proposed approach collects all test cases in $T_n$ that fail on $V_{n-1}$, and notes such test cases as \emph{potential triggering test cases} that may expose the associated bug;
   \item $pt$ is a candidate patch if $V_{n-1}''$ passes \textcolor{black}{at least one} potential triggering test cases. Otherwise, $pt$ is not a valid patch and its validation terminates.
\end{itemize}

If and only if the proposed approach generates exactly a single candidate  patch from a bug-fixing code commit,  the approach recommends the candidate  patch for the associated bug report. Otherwise, no patch would be recommended.

\section{Evaluation}
\label{section:Evaluation}
\subsection{Research Questions}
In this section, we evaluate the proposed approach by investigating the following research questions:
\begin{itemize}[listparindent=-0.5cm,leftmargin=0.3cm,topsep=0.06cm]
\item \textbf{RQ1}:  How often do bug fixing commits contain bug-irrelevant  changes and what percentage of the changes in bug-fixing commits are bug-irreverent?
\item \textbf{RQ2}:  Is BugBuilder accurate in extracting complete and concise bugs/patches? What percentage of real bugs/patches could be extracted accurately by BugBuilder?
\item \textbf{RQ3}:  To what extent does the refactoring detection and replication affect \textcolor{black}{the precision and recall} of BugBuilder?
\item \textbf{RQ4}:  Is BugBuilder scalable?
	\end{itemize}
\subsection{Dataset}
\label{sub:dataset}
Our evaluation is based on the raw data in Defects4J. Defects4J contains 835 real bugs collected from real-world applications. For each bug, it provides the bug-fixing code commit, the versions immediately following/preceding the bug-fixing commit (called $V_n$ and $V_{n-1}$, respectively), and the manually confirmed patch for the bug. $V_n$ and $V_{n-1}$ are taken from version control history by Defects4J without any modification. $V_{n-1}$ is different from the faulty version ($V_{bug}$) provided by Defects4J because $V_{bug}$ is manually constructed to exclude refactorings and feature modifications whereas $V_{n-1}$ is an exact copy from the version control history.

Notably, only $V_n$ and $V_{n-1}$ are leveraged as the input of BugBuilder, whereas the manually constructed patches provided by Defects4J are leveraged only to assess the performance of BugBuilder, i.e., whether the automatically generated patches are identical to the manually constructed ones.  Defects4J contains 835 bug-fixing commits from 17 projects. However, we failed to retrieve the $V_{n-1}$ version for project \emph{Chart} because the version IDs for this project are invalid. Consequently, this project was excluded from our evaluation. As a result, the evaluation was conducted on 809 bug-fixing commits from 16 projects.

\textcolor{black}{Other bug-patch datasets, like iBUGS and ManyBugs, are not leveraged for the evaluation because they do not exclude bug-irrelevant changes from the final bug-fixing patches.}
\subsection{Experiment Design}
\label{sub:experimentDesign}

\subsubsection{RQ1: \textcolor{black}{Prevalence} of Bug-irrelevant  Changes Within Bug-fixing Commits:}
\label{subsub:RQ1-Popularity}
As introduced in Section~\ref{section:Introduction}, bug-irrelevant  changes within bug-fixing commits are preventing us from accepting the whole code commits as patches. To investigate how often bug-fixing commits contain bug-irrelevant  changes, we compared the manually constructed patches in Defects4J against their associated code commits. The comparison was conducted in two steps. First, we investigated how often the patches are identical to   their associated code commits:
  \begin{equation}\label{eq:popularity}
    P_{\rm same}=\frac{\text{\emph{number of commits  identical to associated patches}}}{\text{number of bug-fixing commits}}
  \end{equation}
Assuming that patches in Defects4J are complete and concise, bug-fixing commits that are not identical to the associated patches must contain bug-irrelevant  changes. Consequently,   $P_{\rm diff}=1-P_{\rm same}$ is essentially the percentage of bug-fixing commits that contain bug-irrelevant  changes.

Second, we investigated what percentage of changes in bug-fixing commits are bug-fixing changes and what percentage of changes are bug-irrelevant changes. Because patches in Defects4J have been manually constructed to exclude bug-irrelevant changes~\cite{just2014defects4j},  we took the size of the patches \textcolor{black}{in Defects4J} as the size of bug-fixing changes in the associated bug-fixing commits \textcolor{black}{while the size of the commit is the size of the whole patch in the version-control history}.

\subsubsection{RQ2: Performance of {BugBuilder}}
\label{subsub:RQ2-performance}
To investigate the performance of BugBuilder, we evaluated it on each of the bug-fixing commits in Defects4J as follow:
\begin{itemize}[listparindent=-0.5cm,leftmargin=0.3cm,topsep=0.06cm]
  \item First, we retrieved its associated $V_n$ and $V_{n-1}$ versions as well as the manually constructed patch $pt_{4j}$ associated with the bug-fixing commit;
  \item Second, we leveraged BugBuilder to generate patches, taking $V_{n-1}$ and $V_n$ as input;
  \item Third, if BugBuilder resulted in a patch $pt$, we compared it against the manually constructed patch $pt_{4j}$ to reveal whether the automatically generated patch is identical to the manually constructed one. In case they are identical, we call the generated patch a \emph{matched patch}. \textcolor{black}{Notably, the comparison between generated patches and the ground truth is pure textual comparison of the patches, and it is fully automatic.}
\end{itemize}

An automatically generated patch  is taken as a complete and concise patch if  and only if it is a matched patch, i.e., it is identical to the manually constructed patch (provided by Defects4J) associated with the same bug-fixing commit.

\subsubsection{RQ3: Effect of Refactoring Detection and Reapplication}
\label{subsub:RQ3-design}
As specified in Section~\ref{sub:Refactorings}, BugBuilder excludes refactorings from generated patch by detecting refactorings contained in the bug-fixing commit and reapplying such refactoring to the associated faulty version ($V_{n-1}$).  \textcolor{black}{To investigate to what extent the leveraged refactoring detection and reapplication may affect the precision and recall of BugBuilder,} we disabled refactoring detection and reapplication, and repeated the evaluation (as specified in Section~\ref{subsub:RQ2-performance}).
	

\subsubsection{RQ4: Scalability}
\label{subsub:RQ4-Scalability}
To investigate the scalability of BugBuilder, we depicted the quantitative relation between the run time of BugBuilder and the size of involved  commits.

\subsection{Results and Analysis}
\label{sub:Results}
\subsubsection{RQ1: Bug-irrelevant Changes Are Common in Bug-fixing Commits}
\label{subsub:results:rq1}
\begin{table}
	\small
	\renewcommand{\arraystretch}{1.1}
	\caption{Bug-fixing and bug-irrelevant changes in bug-fixing commits}
	\label{table:PopofBug-IrChanges}
	\centering
	\begin{tabular}{|p{3cm}|c|c|c|}
		\hline
		Project & \tabincell{c}{Size of \\Commits \\($N_1$)} & \tabincell{c}{Size of\\ Bug-fixing\\ Changes ($N_2$)} & \tabincell{c}{$N_{2}/N_{1}$}    \\ \hline
		Commons CLI & 473 & 325 & 69\%  \\
		\hline
		Closure Compiler & 4,108 & 2,111  & 51\%  \\
		\hline
		Commons Codec & 275 &  193  & 70\% \\
		\hline
		Commons Collections & 38 &  29  &  76\%\\
		\hline
		Commons Compress & 602 & 372  & 62\% \\
		\hline
		Commons CSV & 119 &  56  & 47\% \\
		\hline
		Gson & 239 & 168  & 70\% \\
		\hline
		Jackson Core & 485 &  307  & 63\% \\
		\hline
		Jackson Databind & 2,104 &  1,508 & 72\% \\
		\hline
		Jackson Dataformat XML& 119 & 114   & 96\% \\
		\hline
		Jsoup & 1,163 &  777  & 67\% \\
		\hline
		Commons JXPath & 582 &  430  & 74\% \\
		\hline
		Commons Lang & 689 &  516  & 75\% \\
	    \hline
		Commons Math & 1,246 &  763  & 61\% \\
		\hline
		Mockito & 503 &  277  & 55\% \\
		\hline
		Joda Time & 264 &  242  &  92\%\\
		\hline
		 \makecell[c]{TOTAL} & 13,009 &  8,188  & 63\% \\		
		\hline	
	\end{tabular}
\vspace{-0.3cm}
\end{table}

\begin{table*}
\small
\renewcommand{\arraystretch}{1.1}
\caption{Performance of BugBuilder}
\label{performance}
\centering
\begin{tabular}{|l|c|c|c|c|c|}
\hline
   Project & \tabincell{c}{Bug-fixing Commits} & \tabincell{c}{Generated Patches} &  \tabincell{c}{Matched  Patches}& Precision & Recall   \\ \hline

   Commons CLI & 39 & 17 & 17 & 100\% & 44\% \\
   \hline
   Closure Compiler & 174 & 64  & 62 & 97\% & 36\%  \\
   \hline
  Commons Codec & 18 &  10  & 8 & 80\% & 44\% \\
  \hline
  Commons Collections & 4 &  1  &  1 & 100\% & 25\% \\
  \hline
  Commons Compress & 47 & 20 & 19 & 95\% & 40\%  \\
  \hline
  Commons CSV & 16 &  10  & 9 & 90\% & 56\%\\
  \hline
  Gson & 18 & 9   & 8 & 89\% & 44\% \\
  \hline
  Jackson Core & 26 &  10  & 10 & 100\% & 38\% \\
  \hline
  Jackson Databind & 112 &  34  & 33 & 97\% & 29\%\\
  \hline
  Jackson Dataformat XML & 6 & 2   & 2 & 100\% & 33\% \\
  \hline
  Jsoup & 93 &  40  & 36 & 90\% & 39\% \\
  \hline
  Commons JXPath & 22 &  6  & 5 & 83\% & 23\%\\
  \hline
  Commons Lang & 64 &  31  & 29 & 94\% & 45\%\\
  \hline
  Commons Math & 106 &  40  & 39 & 98\% & 37\%\\
  \hline
  Mockito & 38 &  19  & 19 & 100\% & 50\%\\
  \hline
  Joda Time & 26 &  11  & 11 & 100\%& 42\%\\
  \hline
   \makecell[c]{TOTAL} & {809} &  {324}  & {308} & {95\%} & {38\%} \\
 \hline
\end{tabular}
\vspace{-0.3cm}
\end{table*}

For each bug-fixing commit $c$ in Defects4J, we counted the size of the commit (in lines) and the size of bug-fixing changes within the commit. Evaluation results are presented in Table~\ref{table:PopofBug-IrChanges}. From this table, we make the following observations:
\begin{itemize}[listparindent=-0.5cm,leftmargin=0.3cm,topsep=0.06cm]
  \item First, bug-fixing changes account for 63\% of the changes made in bug-fixing commits. In other words, 37\% of the changes in bug-fixing commits are bug-irrelevant, and thus should not be included in patches for the associated bugs. Consequently, taking the whole bug-fixing commits as patches would result in unconcise  patches. Such unconcise patches, if employed by the evaluation of bug-related approaches (e.g., automatic program repair), could be misleading.
  \item Second, the percentage of bug-fixing changes (i.e., the last column of Table~\ref{table:PopofBug-IrChanges}) varies significantly from project to project. It varies from 47\% (on project Commons CSV) to 96\% (on project Jackson Dataformat XML). One possible reason for the variation is that different projects often pose different guidelines on how patches should be committed. 
      However, regardless of the various guidelines posed by different applications, it is unlikely to exclude bug-irrelevant changes completely from all bug-fixing commits.
  \end{itemize}

We also investigated how often  bug-fixing commits contain bug-irrelevant changes, i.e., $P_{\rm diff}$ in Section~\ref{subsub:RQ1-Popularity}. Evaluation results suggest that 379 out of the 809 bug-fixing commits contain bug-irrelevant changes, and thus $P_{\rm diff}=47\%=379/809$. It confirms the conclusion that simply taking the whole bug-fixing commits as patches may frequently (at a chance of 53\%=1-47\%) result in unconcise patches.

From the preceding analysis, we conclude that bug-fixing commits often contain significant bug-irrelevant changes. Consequently, excluding such bug-irrelevant changes from bug-fixing commits is critical for automatic patch extraction.

\subsubsection{RQ2: BugBuilder Is Accurate and Effective}
\label{subsub:results:rq2}
To answer  RQ2, we applied BugBuilder  to each of the bug-fixing code commits in Defects4J and compared its generated patches against the manually constructed patches in Defects4J. If the generated patch is identical to the corresponding patch provided by Defects4J, we call it a \emph{matched patch}.

Evaluation results are presented in Table~\ref{performance}. The first two columns present project names and the number of bug-fixing commits in the projects. The third column presents the number of automatically generated patches. The fourth column presents the number of the generated matched patches, i.e., generated patches that are identical to the manually constructed patches in Defects4J. The last two columns present the precision and recall of BugBuilder.

From this table, we make the following observations:
\begin{itemize}[listparindent=-0.5cm,leftmargin=0.3cm,topsep=0.06cm]
\item First, BugBuilder can generate complete and concise patches on 38\% of the bug-fixing commits. We notice that  BugBuilder  generated 324 patches from 809  bug-fixing commits. We also notice that 308 of the automatically generated patches are identical to manually constructed ones, which results in a recall of 38\%=308/809.
\item Second, BugBuilder is highly accurate. Among the 324 automatically generated patches, 95\%=308/324 are identical to manually constructed ones. We also notice that on 6 out of the 16 projects, BugBuilder achieved 100\% precision, i.e., all patches generated from such projects are both complete and concise.
\end{itemize}

We notice that 16(=324-308) out of the 324 automatically generated patches are different from their corresponding patches in Defects4J. We call them \emph{mismatched patches}. We manually analyzed such mismatched patches, referring to the corresponding patches in Defects4J, associated bug reports, and the associated code commits.
We notice that all of the 16 mismatched patches are supersets of their corresponding patches in Defects4J.
Consequently, the reason for the mismatch should be either (or both) of the following:
\begin{enumerate}
  \item The \textcolor{black}{patches generated by BugBuilder} include some bug-irrelevant changes, i.e., \textcolor{black}{BugBuilder's patches} are complete but not concise;
  \item The manually constructed patches in Defects4J miss some bug-fixing changes,i.e., \textcolor{black}{Defects4J's patches} are incomplete.
\end{enumerate}

\textcolor{black}{Notably, the comparison is based on the specification for concise bug-fixing patches we proposed in Section~\ref{section:Introduction}. Defects4J may have followed different specifications (not explicitly specified in the depository), which could be a potential reason for the difference between the patches in BugBuilder and Defects4J.}

A surprising finding is that the automatically generated patches are often even better than manually constructed patches: 12 out of the 16 mismatched patches are manually confirmed as correct (i.e., both complete and concise) whereas their corresponding patches in Defects4J miss some bug-fixing changes (i.e., incomplete). Counting in such 12 complete and concise patches, the precision of BugBuilder increases to 99\%=(308+12)/324, and its recall increases to 40\%=(308+12)/809. Its precision is even higher than the experts who manually constructed the Defects4J patches: On the 324 commits where BugBuilder generates patches, BugBuilder generates only 4 unconcise patches whereas experts resulted in 12 incomplete patches.

\begin{figure}
	\center
	\includegraphics[width=75mm]{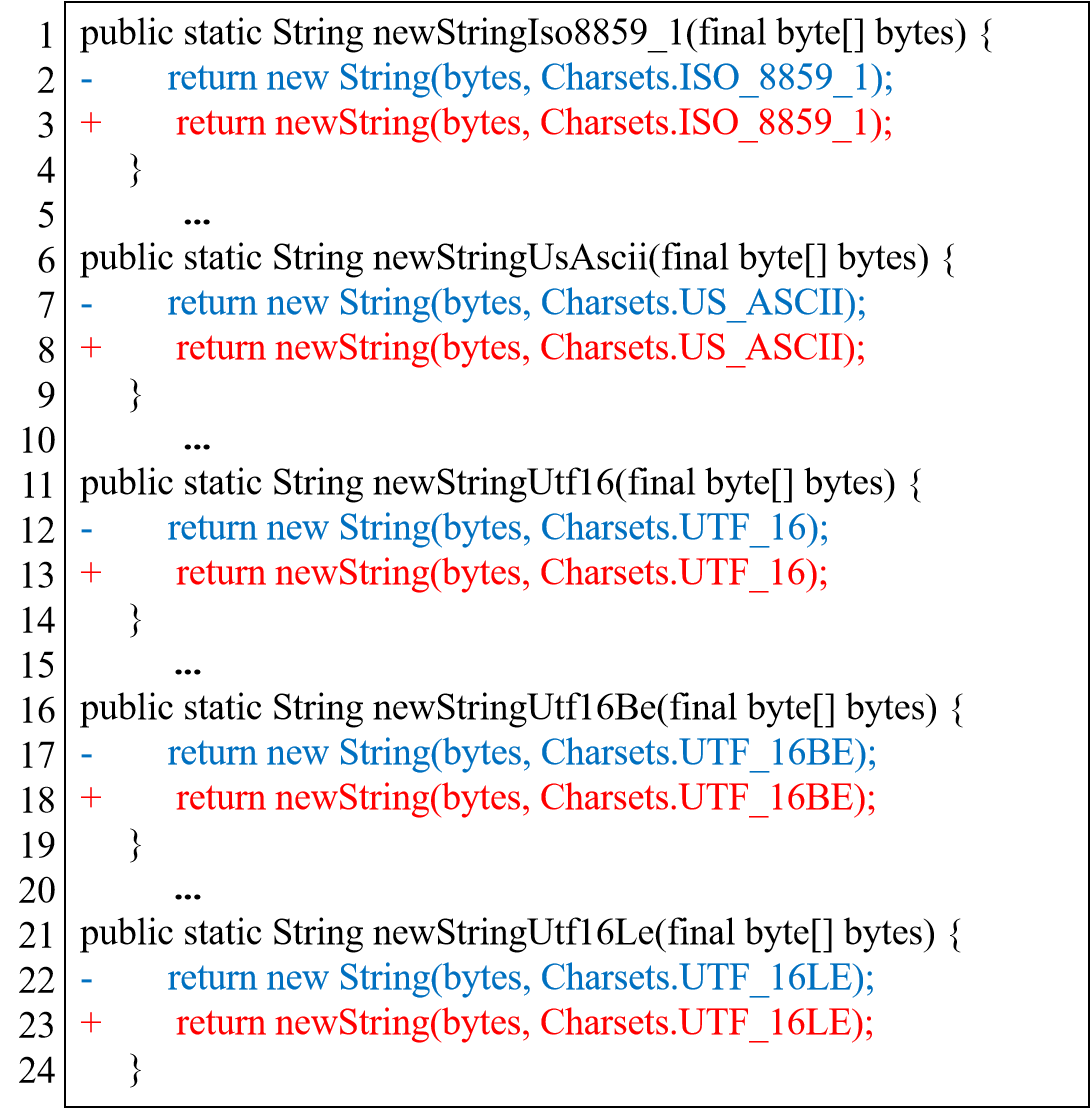}
	\caption{Duplicate Changes in Multiple Places Ignored by Human Experts}
	\label{fig:MulptiplePlaces}
\vspace{-0.4cm}
\end{figure}

A major reason for incomplete patches in Defects4J is that fixing a bug may require duplicate (or highly similar) changes in multiple places (e.g., multiple documents) whereas human experts missed  some places. A typical example is presented in Fig.~\ref{fig:MulptiplePlaces}. This is a bug-fixing commit from project Apache Commons Codec\cite{codec} whose associated bug report is available at\cite{codecbug}. 
As the bug report explains, the return statements  \emph{return new String(bytes, Charsets.xxx)} in a sequence of \emph{newStringxxx} methods (Lines 2, 7, 12, 17, and 22) could not handle null input, and thus they should be replaced with
\emph{return newString(bytes, Charsets.xxx)}. However, the patch in Defects4J \cite{codecbug17} 
contains only the changes in one of the methods (i.e., the first method in Fig.~\ref{fig:MulptiplePlaces}), and thus it is incomplete. In contrast, our approach successfully generates the complete patch containing all of the similar changes in all \emph{newStringxxx} methods.

\begin{figure}
	\center
	\includegraphics[width=90mm]{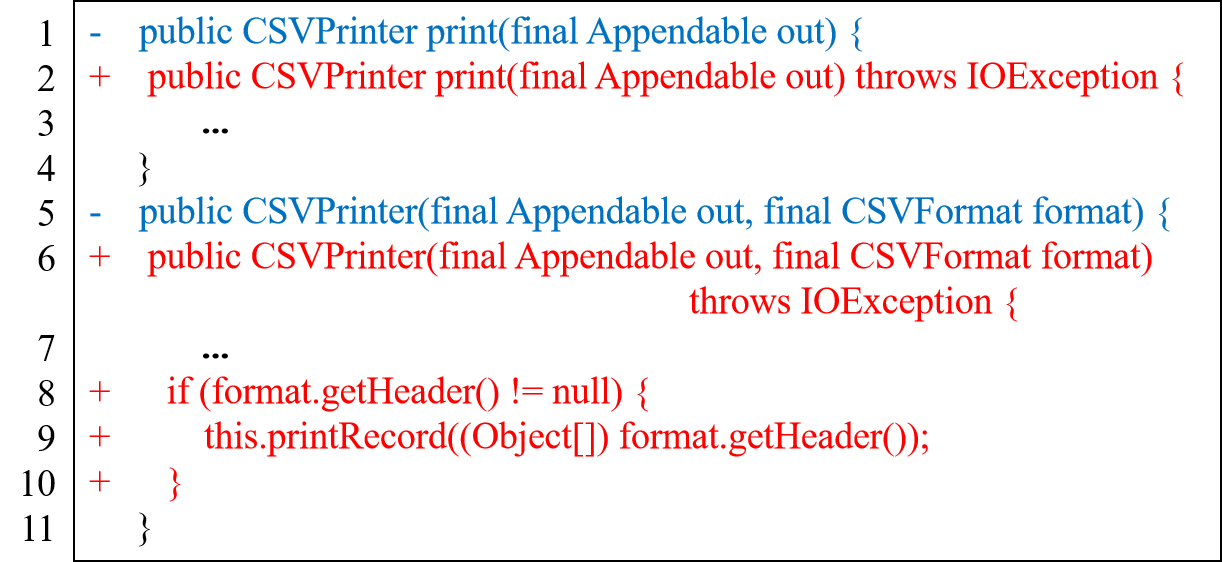}
	\caption{Throw Statements Ignored by Human Experts}
	\label{fig:throwException}
\vspace{-0.4cm}
\end{figure}
The second reason for the incomplete patches in Defects4J is that they ignore the required changes in method declarations and/or variable declarations. A typical example is presented in Fig.~\ref{fig:throwException}. The associated commit comes from Apache Commons CSV\cite{csv} whose associated bug report is available at \cite{csvEBR} 
and its manually constructed patch is available at \cite{csvBug10}. 
As the bug report explains, \emph{CSVFormat} with header does not work with \emph{CSVPrinter}. To fix the bug, the developers added the whole \emph{if} statement (Lines 8-10) to print the header if it is not null.
Notably, the method declaration of \emph{printRecord} explicitly specifies that it has the potential to throw \emph{IOException}.  Consequently, inserting an invocation of this method (Line 9) forces the enclosing method (and its caller, method \emph{print} on Line 1)  to explicitly specify the \emph{IOException} in their method declarations (Line 6 and Line 2).  Otherwise, the revision would result in compiler errors. However, the patch in Defects4J ignores such changes in method declarations, and thus it is incomplete. In contrast, our approach generated the complete patch including the changes in method declarations.

\begin{figure}
	\center
	\includegraphics[width=85mm]{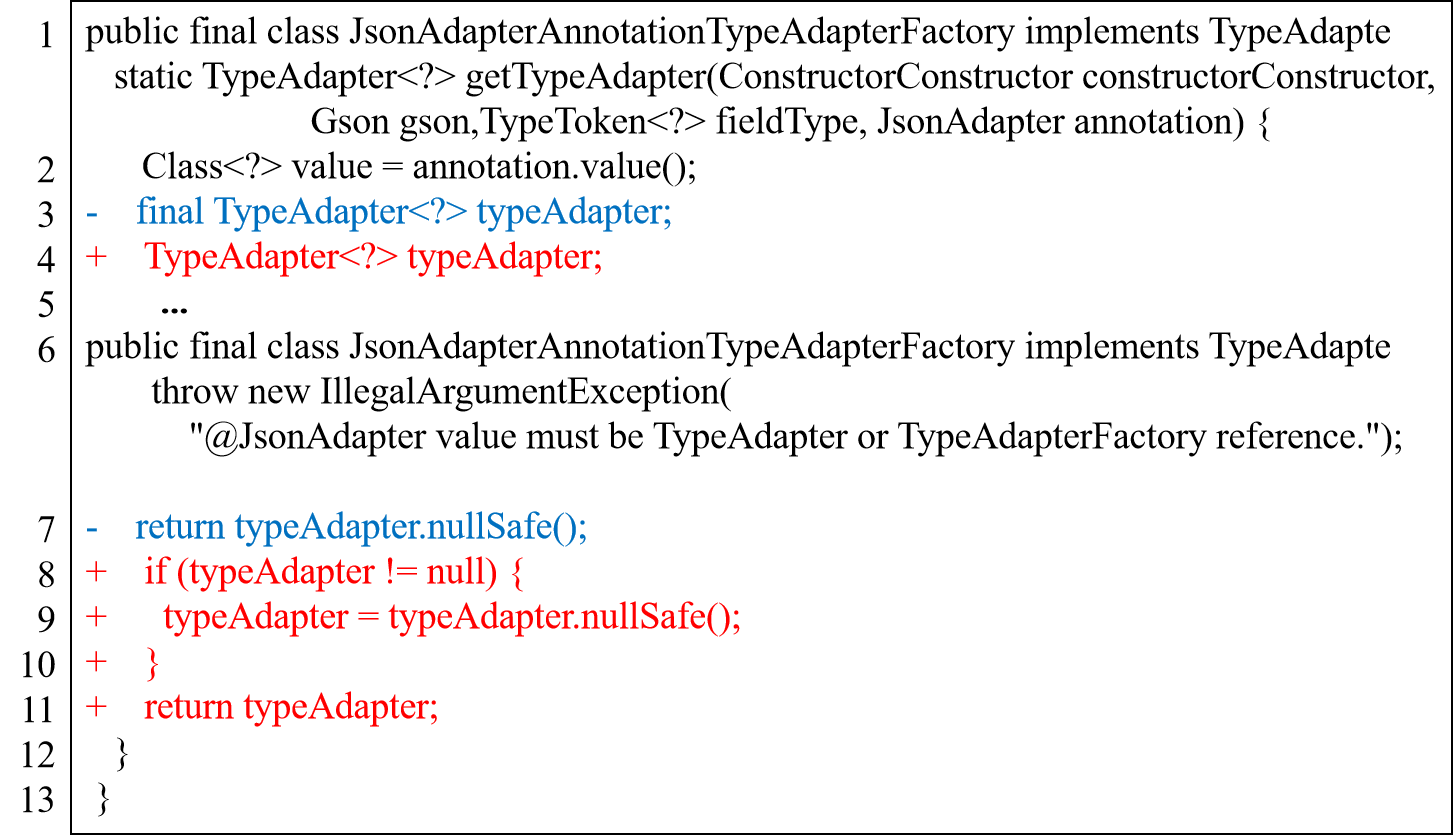}
	\caption{Imperfect Patch Caused by Undiscovered Refactorings}
	\label{fig:IgnoredRefactoring}
\vspace{-0.4cm}
\end{figure}

We also analyzed the four unconcise patches generated by our approach. Analysis results  suggest that all of the 4 unconcise patches are created because the leveraged refactoring-mining tool missed some refactorings  in the involved code commits, and thus such uncovered refactorings were taken as part of the bug-fixing patches.
A typical example is presented in Fig.~\ref{fig:IgnoredRefactoring}.
This example comes from Google Gson\cite{gson}, and the associated bug report is publicly available at \cite{gsonBR}. 
The bug report complains that the method (more specially, the \emph{return} statement on Line 7) would result in null pointer exceptions when \emph{typeAdapter} is null. To fix the bug, developers inserted an \emph{if} statement (Line 8) to validate that \emph{typeAdapter} is not null. The patch provided by Defects4J \cite{gsonBR6} 
is composed of two changes only: Line 8 and Line 10. Other changes are ignored. In contrast, our approach takes all of the changes on the figure as bug-fixing changes. One possible rationale for Defects4J to exclude other changes from the patch is that they could be taken as refactorings: decomposing statement \emph{return typeAdapter.nullSafe();} (Line 7) into two statements \emph{typeAdapter=typeAdapter.nullSafe();} on Line 9 and \emph{return typeAdapter} on Line 11. Because variable \emph{typeAdapter} would not be used anywhere after the \emph{return} statement (Line 11), it could be used as a temporary variable safely. As a result of the usage, the keyword \emph{final} (Line 3) should be removed from the declaration of variable \emph{typeAdapter} because it is assigned/changed on Line 9 as a temporary  variable. We will not argue that such changes should not be taken as refactorings. However, it is a rather complex and unusual \emph{extract variable} refactoring (if it is) because an \emph{extract variable} refactoring usually defines a new variable instead of employing an existing variable temporarily.  Such an unusual refactoring is far beyond the capability of the state-of-the-art refactoring mining tools. Consequently, BugBuilder failed to recognize (let alone reapplying) this refactoring and thus took all of the changes as bug-fixing changes.

\begin{figure}
	\center
	\includegraphics[width=80mm]{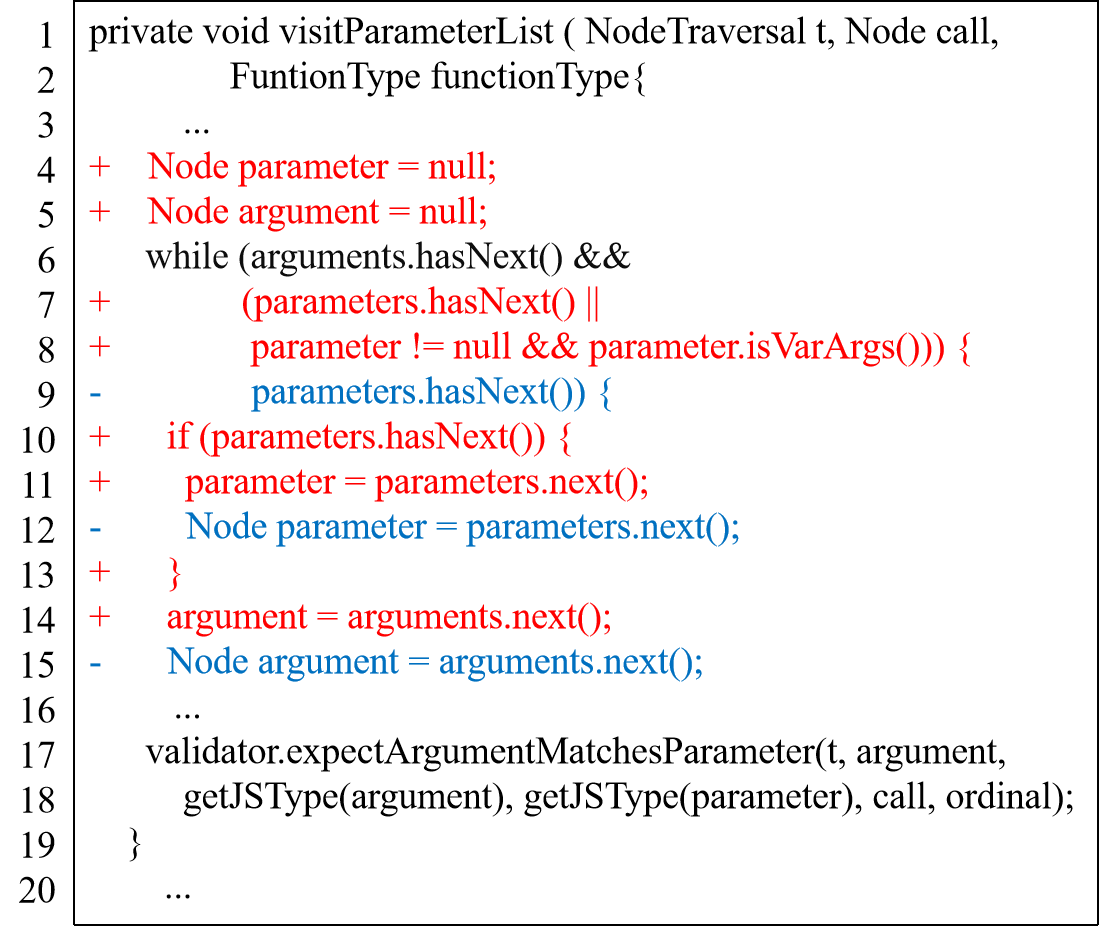}
	\caption{Imperfect Patch Caused by Unsupported Refactorings}
	\label{fig:UnsupportedRefactoring}
\vspace{-0.4cm}
\end{figure}

The example in Fig.~\ref{fig:IgnoredRefactoring} illustrates how unusual refactorings affect BugBuilder, whereas the following example in Fig.~\ref{fig:UnsupportedRefactoring} illustrates how BugBuilder is affected by unsupported refactorings. The bug-fixing commit in Fig.~\ref{fig:UnsupportedRefactoring} comes from Google Closure Compiler\cite{closure} and the associated bug report is available at \cite{closureBR}. 
The bug-fixing changes include the changes on the \emph{while} condition (Lines 7-9) and the \emph{if} statement (Line 10). Such changes are included in both the automatically generated patch and the manually constructed Defects4J patch. However, other changes, i.e., moving the declaration of local variables (\emph{parameter} and \emph{argument}) from the interior of the \emph{while} iteration  (Lines 12 and 15) to  the outside  of the \emph{while} iteration (Lines 4-5), are not taken by Defects4J as bug-fixing changes because they should be taken as  refactorings: The movement would not change the functionality of the method but improves its performance by avoiding the repeating definition of the same variables. However, this kind of refactorings is not yet supported by the refactoring mining tool that is leveraged by BugBuilder. Consequently, BugBuilder failed to remove such refactorings from its generated patches. Notably, it remains controversial whether the refactorings should be excluded from the patch because without such refactorings it is impossible to use variable \emph{parameter} in the \emph{while} condition (as the patch does). However, in this paper, we conservatively take it as a false positive of BugBuilder to avoid controversies.

We conclude based on the preceding analysis that BugBuilder  can generate patches on a significant part (40\%=324/809) of bug-fixing commits in real-world applications, and the generated patches are highly accurate with an average precision of 99\%=(308+12)/324. BugBuilder is even more accurate than human experts. On 324 bug-fixing commits on which BugBuilder generates patches, BugBuilder resulted in 4 unconcise patches whereas human experts resulted in 12 incomplete patches that miss some bug-fixing changes.

\subsubsection{RQ3: Refactoring Detection and Reapplication Improves Recall by 11\%}
\label{subsub:RQ3-results}

\begin{figure}
	\center
	\includegraphics[width=70mm]{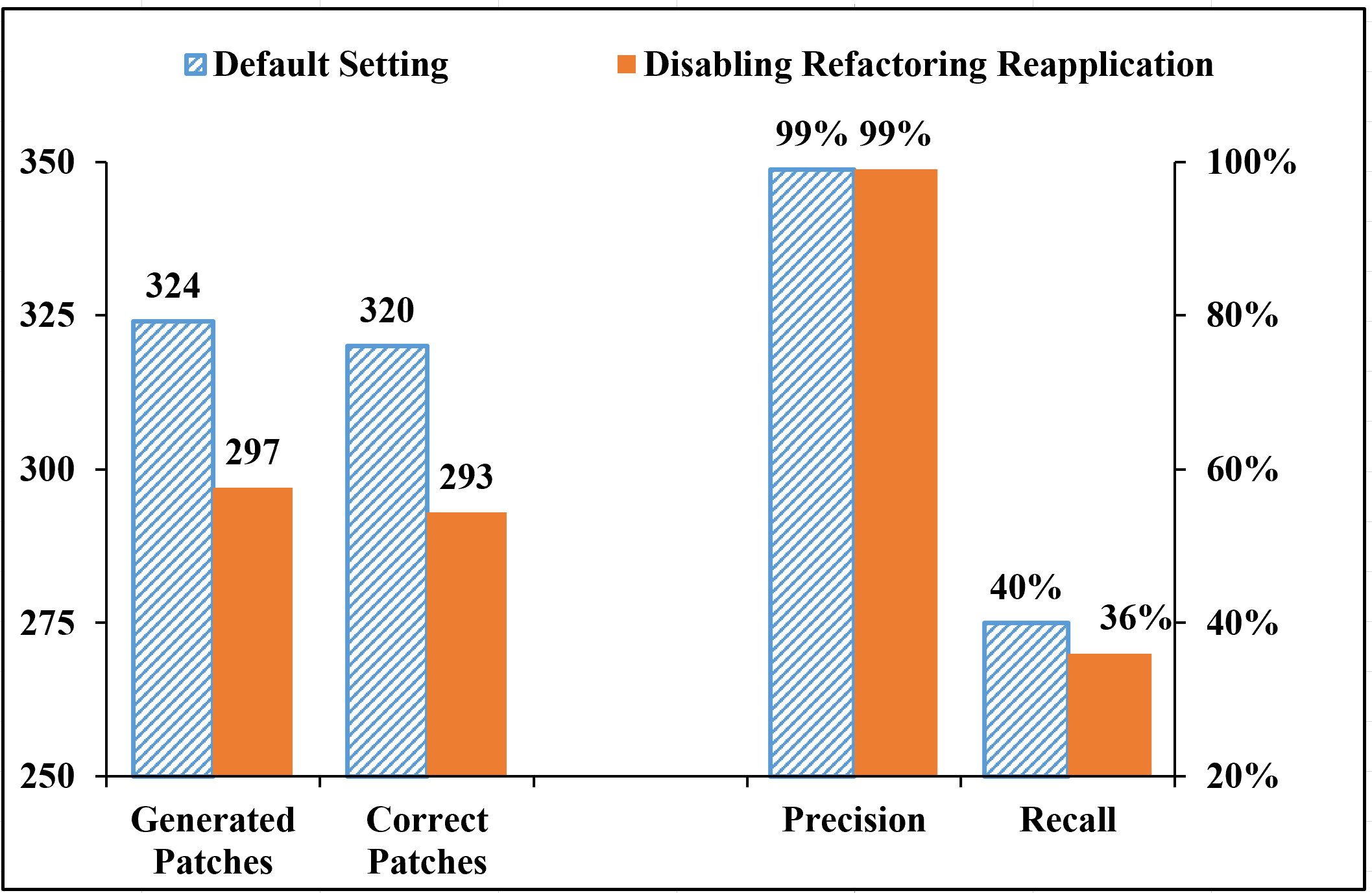}
	\caption{Impact of Refactoring Detection and Replication}
	\label{fig:ImpactOfRefactoring}
\end{figure}

To answer RQ3, we disabled \emph{refactoring detection and reapplication}  in BugBuilder and repeated the evaluation on bug-fixing commits collected by Defects4J. Evaluation results are presented in Fig.~\ref{fig:ImpactOfRefactoring} where \emph{default setting} enables refactoring detection and reapplication. From this  figure, we make the following observations:
\begin{itemize}[listparindent=-0.5cm,leftmargin=0.3cm,topsep=0.06cm]
\item First, enabling or disabling refactoring detection and reapplication has significant impact on the recall of BugBuilder. Enabling it improves the recall of BugBuilder from 36\% to 40\%, resulting in a significant increase of 11\%=(40\%-36\%)/36\%.
\item Second, enabling or disabling refactoring detection and reapplication has no significant impact on the precision of BugBuilder. The precision keeps stable (99\%) regardless of the changes in the setting. \textcolor{black}{A possible reason for the stable precision is that the last part (Section~\ref{sub:PatchGeneration} and Section~\ref{sub:patchValidation}) of the approach can exclude incomplete/unconcise patches, which guarantees high precision regardless of the output of the first part (refactoring-mining) of the approach.}
\end{itemize}

Enabling the detection and reapplication of refactoring increases the number of correctly generated patches (and thus improves the recall) because  bug-fixing commits often contain refactorings. We successfully discover refactorings from 192 out of the involved 809 bug-fixing commits, which we call \emph{refactoring-containing commits}. We also notice that 83 out of the 192 refactoring-containing commits contain no refactorings except for the \emph{supported refactorings} that the current implementation of our approach can identity and reapply. From such commits, BugBuilder (with default setting) successfully generates 27 complete and concise patches. Disabling the detection and reapplication of refactoring, however, fails to generate such patches. Notably, if the implementation of BugBuilder can support additional categories of refactorings in the future, the recall of BugBuilder could be further improved.

An intuitive baseline approach (noted IBA) is that IBA takes the whole bug-fixing commit as a patch if the  commit does not contain any refactorings (not limited to the eight categories of refactorings supported by the current implementation of BugBuilder). In such a way, IBA would generate 617 patches from 809 bug-fixing commits in Defects4J. However 217 out the 617 patches are unconcise, i.e., containing bug-irrelevant changes. Consequently, its precision 65\%=1-217/617 is significantly lower than that (99\%) of BugBuilder.

We conclude based on the preceding analysis that detecting and reapplying refactorings help to improve recall of BugBuilder whereas its precision keeps stable.
%
%
\subsubsection{RQ4: Scalability}
\label{subsub:RQ4-results}

\begin{figure}
	\center
	\includegraphics[width=90mm]{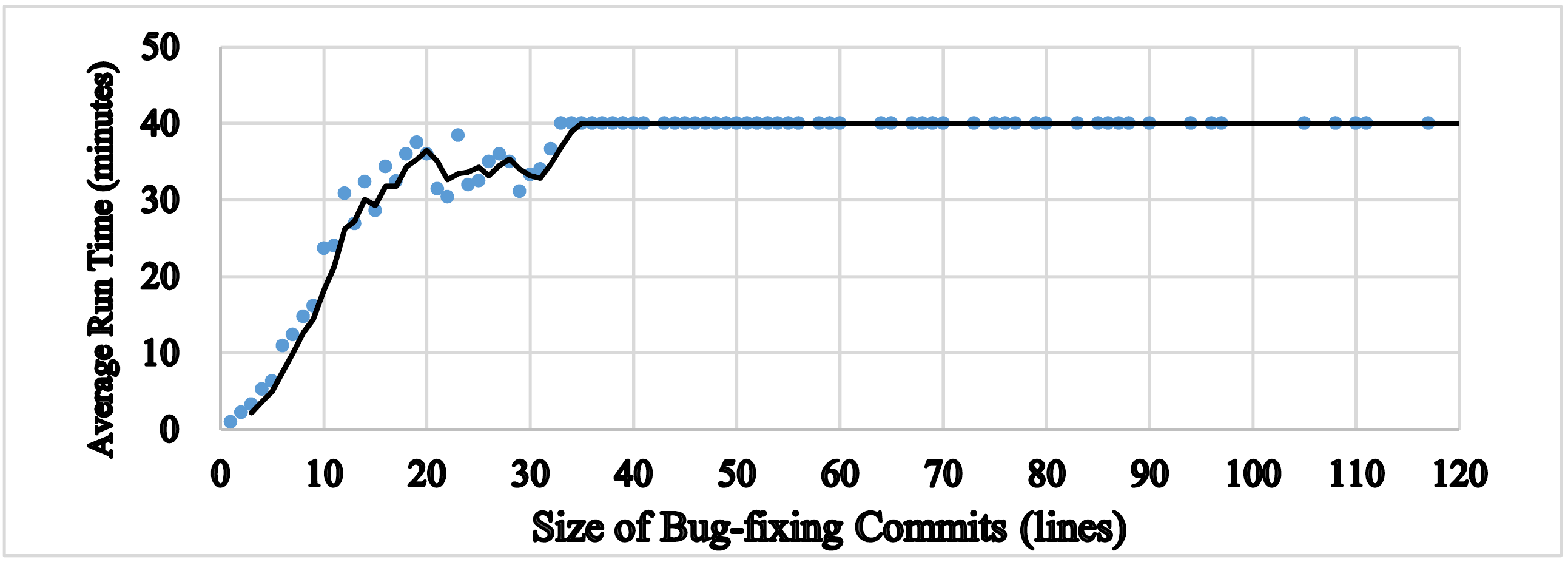}
	\caption{Commit Size Influences Average Run Time}
	\label{fig:scalability}
\end{figure}

Fig.~\ref{fig:scalability} presents a scatter diagram with a trendline, depicting the relationship between the size of bug-fixing commits and the run time of  BugBuilder on such commits. Notably, BugBuilder terminates when its run time reaches 40 minutes on a single commit. From Fig.~\ref{fig:scalability}, we observe that the run time increases significantly with the increase in commit size. We also notice that BugBuilder can efficiently handle bug-fixing commits that contain up to thirty lines of changes. By increasing the maximal time slot (40 minutes at present) for each commit, BugBuilder can even handle larger commits in the future. Currently around 34\% of the commits ran out of the maximal time slot.

Detecting and reapplying refactorings improves the efficiency of BugBuilder. BugBuilder has reapplied refactorings on 83 commits, and its average run time on such commits was 21 minutes.
We disabled the detection and reapplication of refactorings, and reapplied BugBuilder to such commits. Our evaluation results suggest that disabling the detection and reapplication of refactorings increased the average run time on such commits significantly by 24\%=(26-21)/21.

We conclude based on the preceding analysis that BugBuilder is scalable, and most of the commits could be handled within 40 minutes.
\subsection{Threats To Validity}
\label{sub:threats}
The primary threat to external validity is the limited number of involved bug-fixing commits. In the evaluation, we evaluated BugBuilder on 809  bug-fixing commits collected by Defects4J. Special characteristics of such commits may bias the conclusions of the evaluation. We selected such commits for evaluation because Defects4J provides manually constructed concise patches that exclude bug-irrelevant changes. As a result, we can leverage such patches as the ground truth to evaluate the quality of automatically generated patches. To the best of our knowledge, Defects4J is the only bug repository that provides manually constructed concise patches for real bugs in open-source applications, and thus the evaluation was confined to the bug-fixing commits in Defects4J. However, to reduce the threat, we should evaluate BugBuilder on more bug-fixing commits in the future.

A threat to construct validity is that the manual checking of the generated patches (and patches in Defects4J) could be inaccurate. During the evaluation, we manually checked the generated patches and their corresponding patches in Defects4J when they did not match each other. Such manual checking could be biased and inaccurate. To reduce the threat, we presented typical examples in Section~\ref{subsub:results:rq2}, and made all of the generated patches publicly available at~\cite{Patches}.

Another threat to construct validity is that the evaluation is based on the assumption that matched patches are complete and concise. If an automatically generated patch  is identical to the manually constructed patch (in Defects4J) for the same commit, we simply assumed that they are complete and concise. However, as discussed in Section~\ref{subsub:results:rq2}, human experts may also make incorrect (especially incomplete) patches occasionally, and thus the assumption may not always hold.

\section{Discussion}
\label{section:Discussion}

\subsection{It is Critical to Detect and Reapply Refactorings }
A bug-fixing commit may contain three categories of changes: Bug-fixing changes, refactorings, and functionality-related changes (e.g., implementation of new features).  BugBuilder can generate patches for pure bug-fixing commits (without any bug-irrelevant changes) with the potential patch generation (Section~\ref{sub:PatchGeneration}) and validation (Section~\ref{sub:patchValidation}). If a bug-fixing commit contains both bug-fixing changes and refactorings, BugBuilder leverages the \underline{R}efactoring \underline{D}etection and \underline{R}eapplication (RDR for short) to turn the commit into a pure bug-fixing commit, and then generates patches from it. Notably, 24\%=192/809 of the bug-fixing commits in Defects4J contain refactorings, which quantitatively suggests the importance of RDR (Section~\ref{sub:Refactorings}). RDR has the potential to rescue such commits. However, because the current implementation of BugBuilder supports only eight categories of refactorings, which prevents RDR from reaching its maximal potential: It improved recall by 11\% only in the evaluation.

Another significant benefit of RDR is the significant improvement on the BugBuilder's efficiency. Excluding refactorings significantly reduces the size of commits, and thus reduces the number of potential patches. Evaluation results in Section~\ref{subsub:RQ4-results} suggest that disabling RDR  increased  BugBuilder's run time by 24\% on refactoring-contained commits.

\subsection{Extremely High Precision VS Fairish Recall}
High precision (99\%) of BugBuilder is critical for the success of its future application. We expect it to build high-quality bug repositories without any human intervention, and thus the generated patches should be comparable to manually constructed ones. Otherwise, the resulting bug repositories could be misleading and may bias research and evaluation depending on them. Evaluation results in Section~\ref{section:Evaluation} confirm that the automatically generated patches are comparable to (and sometimes better than) patches manually constructed by human experts. In contrast, a fairish recall (40\%) is acceptable because it could be remedied by applying BugBuilder to massive bug-fixing commits outside Defects4J.

\subsection{Limitations}
\textcolor{black}{BugBuilder succeeds on a substantial part (40\%) of the bug-fixing commits, but fails on around 60\% of the commits as well. BugBuilder works on a bug-fixing commit only if 1) the commit is composed of bug-fixing changes only or 2) the commit is composed of only bug-fixing changes and refactorings. Notably, if the refactorings within the commit are only applicable after the bug-fixing changes, existing refactoring mining tools like \emph{RefactoringMiner}~\cite{RefactoringMinerICSE} cannot identify such refactorings by comparing the bug-fixed version ($v_{n}$) and the original buggy version ($v_{n-1}$). For example, if developers insert a fragment of source code to fix a bug, and then apply \emph{extract method} refactoring to extract the inserted source code as a new method, \emph{RefactoringMiner} cannot identify the extract method refactoring because the extracted source code is not available in the original buggy version. As a result, BugBuilder would fail to split the commit accurately into a refactoring patch and a following bug-fixing patch. If the refactorings are required by the bug fix (and thus applied before the fix) or independent of the fix (and thus could be applied before the fix), BugBuilder has the potential to split the bug-fixing commit into a refactoring patch and its following bug-fixing patch. If the commit contains bug-irrelevant and non-refactoring changes (e.g., implementing new features), BugBuilder cannot work either.} Take the bug-fixing commit in Fig.~\ref{fig:throwException} as an example. Adding the functionality to print headers of \emph{CSVFormat} is taken as a bug-fixing action there. However, it could be taken as an implementation of a new feature (printing herders of \emph{CSVFormat}) as well if this functionality has not been specified in the original requirements. Consequently, it is challenging (even for human experts) to distinguish bug-fixing changes from other functionality-related changes without the help of requirements and bug reports. However, automatic and accurate comprehension of requirements and bug reports in plain texts remains challenging, let alone requirements are often unavailable. \textcolor{black}{Most of the bug-fixing commits where BugBuilder failed contain functionality-related bug-irrelevant changes, and this is the major reason for the low recall of BugBuilder.}

\textcolor{black}{Although BugBuilder misses 60\% of the bugs/patches, it enables automatic construction of large bug-patch repositories for the following reasons. First, BugBuilder is fully automated with extremely high precision. Second, BugBuilder is not biased by the types of bugs, but affected by only whether the fixes are mixed with other functionality-related changes. Finally, although Defects4J extracts bug-patches from only 17 projects, there are numerous open-source projects to be exploited. Applying BugBuilder to such projects automatically could significantly increase the capacity of bug-patch repositories, thus offsetting BugBuilder’s weakness (low recall).}

\subsection{Further Improvement on Recall}
In theory, BugBuilder should be able to generate complete and concise patches for all pure bug-fixing commits (without any bug-irrelevant changes). However, BugBuilder succeeded on only 281 out of the 400 pure bug-fixing commits (called \emph{pure commits} for short) in Defects4J. The major reason for the failure is the setting of the maximal time slots: 71\%=84/119 of the failed pure commits ran out of the maximal time slots. Increasing the time slots may improve the recall in the future. Another reason for the failure is the redundancy of some patches. For example, rejecting some changes (i.e., changes on Lines 11-12) of the Defects4J patch  (publicly available at \url{https://github.com/rjust/defects4j/blob/master/framework/projects/Lang/patches/52.src.patch}, but not presented in the paper for space limitation) would not change the semantics of the program. Consequently, BugBuilder generated multiple candidate patches from it, and thus BugBuilder did not known which one should be recommended.

Improving the implementation of the proposed approach to support additional categories of refactorings  may also significantly improve recall in future. Notably, 57\%=109/192 of the refactoring-containing commits in Defects4J contain some refactorings unsupported by the current implementation. Consequently, supporting all such refactoring in future has the potential to double the effect of RDR (\underline{R}efactoring \underline{D}etection and \underline{R}eapplication) that currently improves recall by 11\%.

\section{Conclusions and Future Work}
\label{section:conclusions}
Large-scale and high-quality repositories of real bugs are critical for bug-related research. However, existing approaches to building such repositories either fail to exclude bug-irrelevant changes from patches or require human intervention. To this end, in this paper, we propose a novel approach, called \emph{BugBuilder}, to extracting complete and concise patches from bug-fixing commits automatically. 
BugBuilder has been evaluated on 809 bug-fixing commits in Defects4J. Evaluation results suggest that it successfully generates complete and concise patches for  forty percent of the bug-fixing commits, and its precision was even higher than human experts.

With the help of BugBuilder, we plan to build large-scale and high-quality repositories of real bugs automatically in the future. It is also practical and meaningful to build repositories of bugs in specific domains or planforms, e.g., real bugs in mobile applications and real bugs in machine learning applications. Automatic approaches (e.g., BugBuilder) to generating complete and concise patches from bug-fixing commits may make such challenging tasks practical.

\section*{Acknowledgments}
This work was sponsored in part  by the National Key Research and Development Program of  China (2017YFB1001803), and the National Natural Science  Foundation of China (61690205, 61772071).

\bibliographystyle{IEEEtran}
\bibliography{IEEEabrv,reference}
\end{document}